\documentstyle[12pt]{article}
\titlepage

\title{Cohomology and MP Spacetimes}
\author{  Richard Atkins \\
        richard.atkins@twu.ca \\
        Department of Mathematics\\ 
		Trinity Western University \\
		7600 Glover Road \\
		Langley, BC, V2Y 1Y1 Canada}
\date{}
\newtheorem{fact}{Fact}

\newtheorem{lemma}[fact]{Lemma}
\newtheorem{proposition}[fact]{Proposition}
\newtheorem{theorem}[fact]{Theorem}
\newtheorem{corollary}[fact]{Corollary}
\begin{document}
\maketitle
\begin{abstract}
This paper pursues a cohomological formulation for gravitation in which gravity might be
expressed in terms of a gravitational potential, much in the spirit of electrodynamics. 
To this end we introduce a cochain complex consisting of $(n+2)$-tensors, symmetric in two
indices and skew-symmetric in the remaining $n$ indices. The cohomology of the complex 
is shown to be isomorphic to the \v{C}ech cohomology of an appropriately defined sheaf
of functions. Furthermore, it is demonstrated that in isotropic coordinates MP spacetimes 
as well as the Schwarzschild black hole may be thusly represented, that is, by means of 
the coboundary of a potential, as defined by the differential operator of the complex.
\end{abstract}

\newpage

\section{Introduction}
The field strength in classical electrodynamics is given by a symmetric-free two-tensor
$F=F_{\mu \nu}$, defining an element in the second de Rham cohomology group. 
The motion of a charged particle is then determined by
\begin{equation} \label{electro} 
\ddot{x}^{\mu} = -kF^{\mu}_{\nu} \dot{x}^{\nu}
\end{equation} 
for some constant $k$, with respect to a flat background structure. 
The associated quantum field theory rests upon this cohomological foundation in that
the dynamical variables to be quantized are the electrodynamic potentials, whose coboundaries
in the de Rham complex give the field strength.  
This leads us to surmise whether a quantum description of gravitation demands a similar 
representation for an appropriately chosen cohomology theory defined, at least,
for a significant class of spacetimes. 

A free particle in a gravitational field follows a geodesic:
\[ \ddot{x}^{\mu} = -\Gamma^{\mu}_{\nu \lambda}\dot{x}^{\nu}\dot{x}^{\lambda} \]
The form of these equations of motion suggests, by analogy to equation (\ref{electro}),
the interpretation of the connection as the gravitational field strength. 
We introduce a flat metric $\eta=diag(+1,-1,-1,-1)$ with respect to
the coordinates $x^{\mu}$ and define 
$\Gamma_{\mu \nu \lambda} = \eta_{\mu \tau}\Gamma^{\tau}_{\nu \lambda}$.
The connection can be uniquely decomposed into a sum of a 
completely symmetric part $\Gamma_{(\mu \nu \lambda)}$ and a symmetric-free part
$F_{\mu \nu \lambda}$
\[  \Gamma_{\mu \nu \lambda} = \Gamma_{(\mu \nu \lambda)} + F_{\mu \nu \lambda} \]
This paper develops a cohomology theory for symmetric-free tensors of the form  
\[ S_{\mu_{1}\cdots\mu_{n}\nu_{1}\nu_{2}}\] which are skew-symmetric in the $\mu$ indices and 
symmetric in the $\nu$ indices. It is shown that in isotropic coordinates,
the symmetric-free part of the field strength of Majumdar-Papapetrou
(MP) spacetimes (cf. \cite{dd}, \cite{kk}, \cite{ll}, \cite{mm}) is the coboundary of a 
potential in the associated cochain complex. Since $\Gamma^{\mu}_{\nu \lambda}$ is 
not a tensorial object the existence of such potentials is dependent upon the 
choice of coordinates considered; indeed, such a characterization of the connection will hold 
only in a very restricted class of coordinate systems, if at all. Thus this formulation
introduces a kind of gauge fixation of the diffeomorphism invariance inherent in 
general relativity.

In the following section a cochain complex $(K^{*}(M),d_{K})$ is defined for manifolds  $M$ 
endowed with a flat connection (cf.  \cite{aa}, \cite{bb}, \cite{tt}). 
The associated cohomology groups $H^{q}_{K}(\Re^{n})$ for 
Euclidean space is determined by application of the Poincar\'{e} Lemma. 
The cohomology $H^{*}_{K}(M)$ for a general manifold $M$ is then identified with the \v{C}ech 
cohomology of the sheaf of local affine functions using spectral sequences
(cf. \cite{cc}, \cite{hh}, \cite{ii}). 
Section 3 defines a second cochain complex $(G^{*}(M),d_{G})$ isomorphic to
$(K^{*}(M),d_{K})$ which relates more directly to gravitation. In the final section,
we consider the MP class of spacetime solutions along with the
Schwarzschild black hole and show that they may be written in terms of a gravitational 
potential in the manner described above.

\section{Cohomology of the $K^{*}(M)$  complex}

In this section we define a cochain complex $(K^{*}(M), d_{K})$, which will be intermediary to 
the cohomology theory associated to gravitation, to be developed in the next section. 
We begin with some basic definitions and conventions.

Let $M$ be a manifold and $\nabla$ a
flat connection on $M$. $\nabla$ acts on a covariant $n$-tensor 
$C= C_{\mu_{1}\cdots\mu_{n}}$ on $M$ by introducing an index to the left:
\[ (\nabla C)_{\mu_{1}\cdots\mu_{n+1}}dx^{\mu_{1}}\otimes \cdots \otimes dx^{\mu_{n+1}} = 
dx^{\mu_{1}} \otimes \nabla_{\frac{\partial}{\partial x^{\mu_{1}}}}(C) \]
This shall be expressed more conveniently in terms of indices by
\[ (\nabla C)_{\mu_{1}\cdots\mu_{n+1}} = \nabla_{\mu_{1}} C_{\mu_{2}\cdots\mu_{n+1}} \]
Define skew-symmetrization $s(C)$ by
\[ s(C)_{\mu_{1}\cdots\mu_{n}} = C_{[\mu_{1}\cdots\mu_{n}]} = 
        \frac{1}{n!}\sum_{\sigma \in S_{n}}sg(\sigma)C_{\mu_{\sigma(1)}\cdots\mu_{\sigma(n)}} \]
where the sum ranges over all permutations $\sigma$ on $n$ letters, $S_{n}$. 
Thus, $s(s(C))=s(C)$. Define $d_{\nabla}$ by
\[ (d_{\nabla}C)_{\mu_{1}\cdots\mu_{n+1}} = 
 \nabla_{[\mu_{1}}C_{\mu_{2}\cdots\mu_{n}]\mu_{n+1}} \]
Since $\nabla$ is flat,  $d_{\nabla}^{2} = 0$.

Let $Sym^{n}(M)$ denote the symmetric sections of $T^{n,0}M$, the tensor product of $n$ 
copies of the cotangent bundle. The module $K^{n}(M)$ over the 
ring of smooth functions $\Omega^{0}(M)$ on $M$ is defined, for $n \geq 2$, 
to be the submodule of 
 \[ \Omega^{n}(M)\otimes_{\Omega^{0}(M)} Sym^{1}(M) \]
consisting of tensor fields $T_{\mu_{1}\cdots\mu_{n}\nu}$ satisfying the condition
\begin{eqnarray} T_{[\mu_{1}\cdots\mu_{n}\nu]} = 0 \label{K}
\end{eqnarray}
$\Omega^{n}(M)$, as usual, denotes the module of $n$-forms on $M$. 
Set $K^{0}(M) = \Omega^{0}(M)$ and $K^{1}(M)=Sym^{2}(M)$. 
The cochain complex $(K^{*}(M),d_{K})$ is defined to be
\[ 0 \longrightarrow K^{0}(M)  \stackrel{\nabla^{2}}{\longrightarrow} K^{1}(M) 
    \stackrel{d_{\nabla}}\longrightarrow  K^{2}(M) \stackrel{d_{\nabla}}\longrightarrow 
    K^{3}(M) \stackrel{d_{\nabla}}\longrightarrow \cdots   \]
The cohomology of the complex is 
\[ H^{*}_{K}(M) = \frac{ker \hspace {0.03in} d_{K}}{im \hspace{0.03in} d_{K}} \]
which naturally inherits a grading from $K^{*}(M)$.

\begin{theorem} \label{theorem:poincare} 
 \[ H^{q}_{K}(\Re^{n}) = \left\{ 
\begin{array}{ll}
\Re^{n+1} &  \hspace{0.2in}   \mbox{if \hspace{0.03in} $q = 0$} \\
0 &  \hspace{0.2in} \mbox{otherwise} 
\end{array}  \right. \]
\end{theorem}
{\it Proof:}\\
We work in coordinates of $\Re^{n}$ for which covariant
differentiation with respect to $\nabla$ coincides with partial differentiation.
 
Let $q=0$. $f\in K^{0}(\Re^{n})$ is a cocycle iff $\partial_{\mu}\partial_{\nu}f = 0$
for all $\mu$ and $\nu$. In this case, $f$ has the form $f=a_{\mu}x^{\mu} + b$ for some
real constants $a_{\mu}$ and $b$. Therefore $H^{0}_{K}(\Re^{n})=\Re^{n+1}$.

Suppose $q=1$. Let $T=T_{\mu \nu} \in K^{1}(\Re^{n})$ be a cocycle. Then 
$\partial_{\mu}T_{\nu \lambda} =   \partial_{\nu}T_{\mu \lambda}$. By the Poincar\'{e} 
Lemma, there exist functions $f_{\mu}$ defined on $\Re^{n}$ such that 
$T_{\mu \nu} = \partial_{\mu}f_{\nu}$. 
Since $T$ is symmetric, $\partial_{\mu}f_{\nu} = \partial_{\nu}f_{\mu}$. Thus there 
exists $f\in K^{0}(\Re^{n})$ such that $f_{\mu} = \partial_{\mu}f$, by the Poincar\'{e} Lemma again.
Therefore $T_{\mu \nu}= \partial_{\mu}\partial_{\nu}f$ and so $T=d_{K}f$. This shows that 
$H^{1}_{K}(\Re^{n})=0$.

Now consider $q>1$ and suppose $T = T_{\mu_{1}\cdots \mu_{q}\nu} \in K^{q}(\Re^{n})$ is a cocycle:
\[ \partial_{[\mu_{1}}T_{\mu_{2}\cdots\mu_{q+1}]\nu} =0 \] 
By the Poincar\'{e} Lemma, there exist
$(q-1)$-forms $A_{\nu}=A_{\mu_{1}\cdots\mu_{q-1}\nu}$, skew-symmetric in the $\mu$ indices,
such that 
$T_{\mu_{1}\cdots \mu_{q}\nu} = \partial_{[\mu_{1}}A_{\mu_{2}\cdots\mu_{q}]\nu}$. However,
$A=A_{\nu}=A_{\mu_{1}\cdots\mu_{q-1}\nu}$ is not necessarily an element of 
$K^{q-1}(\Re^{n})$ since the condition 
\[ s(A)=A_{[\mu_{1}\cdots\mu_{q-1}\nu]}=0 \]
is not guaranteed. In order to remedy this we 
make use of the freedom available in the choice of $A$. Observe that
\[ ds(A)= s(\nabla A) = s(d_{\nabla}A) = s(T) = 0\]
Therefore, $s(A)=dB$ for some $(q-1)$-form $B$, by applying the Poincar\'{e} Lemma once more. 
Define $A'=A-d_{\nabla}B$. Then
\[ s(A')=s(A)-s(d_{\nabla}B) = s(A)-dB = 0 \]
Furthermore,
\[ d_{K}A' = d_{\nabla}(A)-d_{\nabla}(d_{\nabla}B) = T-d_{\nabla}^{2}B = T \]
Therefore $A'\in K^{q-1}(\Re^{n})$ is a preimage of $T$ under $d_{K}$ and so
$H^{q}_{K}(\Re^{n})=0$. \\
$\Box$ \\

Next, we seek to relate the $K^{*}(M)$ cohomology to the more familiar \v{C}ech cohomology. 
Let {\it Aff} denote the sheaf on $M$ whose sections over an open subset 
$U \subseteq M$ is the kernel of the map 
$\nabla^{2}:\Omega^{0}(U) \longrightarrow Sym^{2}(U)$. That is, {\it Aff} is the sheaf of
local affine functions. ${\cal K}^{q}$ shall denote the sheaf whose sections over 
$U$ is ${\cal K}^{q}(U)=K^{q}(U)$.

\begin{theorem} $H^{n}_{K}(M) \cong H^{n}(M,\mbox{Aff})$, for all $n\geq 0$.
\end{theorem}
{\it Proof:}\\
The demonstration is similar to the spectral sequence argument 
used to prove the \v{C}ech-de Rham isomorphism.

Let ${\cal U}= \{U_{i}: i\in I\}$ be a good cover of $M$.
Consider the double complex $F = \oplus F^{p, q}$, where
\[ F^{p, q} = {\cal C}^{p}({\cal U},{\cal K}^{q}) 
= \prod_{\alpha_{0}< \alpha_{1}<\cdots<\alpha_{p}}
K^{q}(U_{\alpha_{0}\alpha_{1}\cdots\alpha_{p}}) \]
and 
\[ U_{\alpha_{0}\alpha_{1}\cdots\alpha_{p}} = 
U_{\alpha_{0}}\cap U_{\alpha_{1}} \cap \cdots \cap U_{\alpha_{p}} \]
$F$ is equipped with two differential operators, $d_{K}:F^{p,q} \longrightarrow F^{p,q+1}$
and $\delta: F^{p,q} \longrightarrow F^{p+1,q}$, the \v{C}ech coboundary operator defined by
\[ (\delta\omega)_{\alpha_{0}\alpha_{1}\cdots\alpha_{p+1}} =  
    \sum_{i=0}^{p+1}(-1)^{i}
    \omega_{\alpha_{0}\cdots\widehat{\alpha_{i}}\cdots\alpha_{p+1}
    \textstyle{|_{U_{\alpha_{0}\alpha_{1}\cdots\alpha_{p+1}}}}}    \]
for $\omega \in F^{p,q}$. 
Since the ${\cal K}^{q}$ are fine sheaves, the $E'_{1}$ term of the 
second spectral sequence is
\[ E'^{p, q}_{1} = H^{p, q}_{\delta} = \left\{
   \begin{array}{ll}
    K^{q}(M)  &  \hspace{0.2in}   \mbox{if \hspace{0.03in} $p = 0$} \\
0 &  \hspace{0.2in} \mbox{otherwise} 
\end{array}  \right. \]
The $E'_{2}$ term is
\[ E'^{p, q}_{2} = H^{p, q}_{d_{K}}H_{\delta} = \left\{
   \begin{array}{ll}
    H_{K}^{q}(M)  &  \hspace{0.2in}   \mbox{if \hspace{0.03in} $p = 0$} \\
0 &  \hspace{0.2in} \mbox{otherwise} 
\end{array}  \right. \]
Since $E'^{p, q}_{2}=E'^{p, q}_{\infty}$,
\begin{equation} \label{co1}
 H^{n}_{K}(M) \cong H^{n}_{D}(F) 
\end{equation}
where the right hand side is the cohomology of the double complex.

By the previous theorem, the first spectral sequence has $E_{1}$ term 
\[ E^{p, q}_{1} = H^{p, q}_{d_{K}} = \left\{
   \begin{array}{ll}
    {\cal C}^{p}({\cal U},\it{Aff})  &  \hspace{0.2in}   \mbox{if \hspace{0.03in} $q = 0$} \\
0 &  \hspace{0.2in} \mbox{otherwise} 
\end{array}  \right. \]
The $E_{2}$ term is
\[ E^{p, q}_{2} = H^{p, q}_{\delta}H_{d_{K}} = \left\{
   \begin{array}{ll}
    H^{p}({\cal U},\it{Aff})  &  \hspace{0.2in}   \mbox{if \hspace{0.03in} $q = 0$} \\
0 &  \hspace{0.2in} \mbox{otherwise} 
\end{array}  \right. \]
$E^{p, q}_{2}=E^{p, q}_{\infty}$ and so,
\begin{equation} \label{co2}
 H^{n}({\cal U}, \it{Aff}) \cong H^{n}_{D}(F) 
\end{equation}
From the isomorphisms (\ref{co1}) and (\ref{co2}), 
\[ H^{n}_{K}(M) \cong  H^{n}({\cal U}, \it{Aff}) \]
Taking the direct limit of $H^{n}({\cal U}, \it{Aff})$ yields the
desired result.\\
$\Box$ \\

\section{Cohomology of the $G^{*}(M)$ complex}

A second cochain complex $(G^{*}(M), d_{G})$ isomorphic to
$(K^{*}(M), d_{K})$ is introduced, with  which we may give a cohomological
description of certain spacetimes.   

Let $\tau \in S_{n+1}$ be the cyclic permutation 
$(1\hspace{0.07in} 2\hspace{0.07in} 3 \hspace{0.07in} \cdots \hspace{0.07in} n+1)$.
The $ith$ power $\tau^{i}$ of $\tau$ is the cycle given by $\tau^{i}(k) = k+i$
mod $n+1$, where the remainders upon division by $n+1$ are $\{1,...,n+1\}$. For instance,
$\tau^{2}(1)=3, \tau^{2}(2)=4,..., \tau^{2}(n-1)=n+1, \tau^{2}(n)=1$ and $\tau^{2}(n+1)=2$.

For $n\geq 2$, define the $\Omega^{0}(M)$-module $G^{n}(M)$ to be the submodule of
 \[ \Omega^{n-1}(M)\otimes_{\Omega^{0}(M)} Sym^{2}(M) \]
consisting of tensor fields $S_{\mu_{1}\cdots\mu_{n+1}}$ satisfying the condition
\[ \sum_{i=0}^{n} (-1)^{in}S_{\mu_{\tau^{i}(1)}\cdots\mu_{\tau^{i}(n+1)}} = 0 \]
Explicitly, 
\begin{eqnarray} \label{G}
S_{\mu_{1}\cdots\mu_{n+1}}  +  (-1)^{n}S_{\mu_{2}\cdots\mu_{n+1}\mu_{1}}  +  
S_{\mu_{3}\cdots\mu_{n+1}\mu_{1}\mu_{2}} +  
(-1)^{n}S_{\mu_{4}\cdots\mu_{n+1}\mu_{1}\mu_{2} \mu_{3}}+ \nonumber \\
S_{\mu_{5}\cdots\mu_{n+1}\mu_{1}\mu_{2} \mu_{3}\mu_{4}} +
\cdots  + (-1)^{n}S_{\mu_{n+1}\mu_{1}\cdots\mu_{n}} = 0 
\end{eqnarray}

Define $\phi^{n}: K^{n}(M)\longrightarrow G^{n}(M)$ by 
\[ \phi^{n}(T)_{\mu_{1}\cdots\mu_{n-1}\nu \lambda} = T_{\mu_{1}\cdots\mu_{n-1}(\nu\lambda)}
   =\frac{1}{2}
(T_{\mu_{1}\cdots\mu_{n-1}\nu\lambda}+T_{\mu_{1}\cdots\mu_{n-1}\lambda\nu}) \]
for $T = T_{\mu_{1}\cdots\mu_{n-1}\nu\lambda} \in K^{n}(M)$.
Let $\psi^{n}: G^{n}(M)\longrightarrow K^{n}(M)$ be defined by
\[ \psi^{n}(S)_{\mu_{1}\cdots\mu_{n}\nu} = \frac{2n}{n+1}S_{[\mu_{1}\cdots\mu_{n}]\nu} \]
for $S=S_{\mu_{1}\cdots\mu_{n}\nu} \in G^{n}(M)$.
Set $G^{0}(M)=K^{0}(M)= \Omega^{0}(M)$, $G^{1}(M)=K^{1}(M)=Sym^{2}(M)$ and define 
$\phi^{n}: K^{n}(M)\longrightarrow G^{n}(M)$ and $\psi^{n}: G^{n}(M)\longrightarrow K^{n}(M)$
to be the identity for $n=0,1$. Henceforth, we will  drop the superscript "$n$" for these
mappings. In what follows, it shall be shown that $\phi$ and $\psi$ are inverse to each 
other.

\begin{proposition} \label{prop:idK}
$ \psi\circ\phi = Id_{K^{*}} $
\end{proposition}
{\it Proof:}\\
Let $T= T_{\mu_{1}\cdots\mu_{n+1}} \in K^{n}(M)$, for $n\geq 2$. 
Denote $S= \phi(T) \in G^{n}(M)$.
Then 
\begin{eqnarray}
    &   & \psi\circ\phi(T)_{\mu_{1}\cdots\mu_{n+1}} \nonumber \\
    & = &  \psi(S)_{\mu_{1}\cdots\mu_{n+1}}   \nonumber \\
    & = & \frac{2n}{n+1}S_{[\mu_{1}\cdots\mu_{n}]\mu_{n+1}} \nonumber \\
    & = & \frac{2n}{n+1}\frac{1}{n!}\sum_{\sigma \in S_{n}}
    sg(\sigma)S_{\mu_{\sigma(1)}\cdots\mu_{\sigma(n)}\mu_{n+1}} \nonumber \\
    & = & \frac{2n}{n+1}\frac{1}{n!}\sum_{\sigma \in S_{n}}sg(\sigma)\frac{1}{2}
         \left(T_{\mu_{\sigma(1)}\cdots\mu_{\sigma(n)}\mu_{n+1}} +
          T_{\mu_{\sigma(1)}\cdots\mu_{\sigma(n-1)}\mu_{n+1}\mu_{\sigma(n)}}\right) \nonumber \\
    & = &  \frac{n}{n+1}\frac{1}{n!}
         \left(n!T_{\mu_{1}\cdots\mu_{n+1}} +
         \frac{1}{n}P\right) \nonumber 
   \end{eqnarray} 
where
\begin{eqnarray}
P & = & \sum_{i=1}^{n}(-1)^{n-i}\sum_{\sigma \in S_{n}}sg(\sigma)
          T_{\mu_{\sigma(1)}\cdots\mu_{\sigma(i-1)}\mu_{n+1}
         \mu_{\sigma(i)}\cdots\mu_{\sigma(n)}} \nonumber \\
  & = &  -\sum_{\tau\in S_{n+1}} 
          sg ( \tau )T_{\mu_{\tau(1)}\cdots\mu_{\tau(n+1)}}+  
           \sum_{\sigma \in S_{n}}sg ( \sigma )
          T_{\mu_{\sigma(1)}\cdots\mu_{\sigma(n)}\mu_{n+1}}  \nonumber   
\end{eqnarray}           
The last equality follows from the fact that if 
\[(\tau(1),...,\tau(n+1))= (\sigma(1),...,\sigma(i-1),n+1,\sigma(i),...,\sigma(n)) \]
for $\tau\in S_{n+1}$ and $\sigma \in S_{n}$ then 
$sg(\tau)=(-1)^{n-i+1}sg(\sigma) = -(-1)^{n-i}sg(\sigma)$.
Hence
\begin{eqnarray}
P & = & -(n+1)!T_{[\mu_{1}\cdots\mu_{n+1}]} + n!T_{\mu_{1}\cdots\mu_{n+1}}
          \nonumber \\
  & = &   n!T_{\mu_{1}\cdots\mu_{n+1}} \nonumber   
\end{eqnarray} 
by property (\ref{K}) for elements $T\in K^{n}(M)$.
Substituting this into the expression for  $\psi\circ\phi(T)$ obtained above gives
\begin{eqnarray}   
\psi\circ\phi(T)_{\mu_{1}\cdots\mu_{n+1}} & = &  
      \frac{n}{n+1}\frac{1}{n!}
         \left(n!T_{\mu_{1}\cdots\mu_{n+1}} +
         \frac{1}{n} n!T_{\mu_{1}\cdots\mu_{n+1}}\right) \nonumber \\
  & = &  T_{\mu_{1}\cdots\mu_{n+1}} \nonumber
\end{eqnarray}  
$\Box$ \\

\begin{lemma} \label{lemma:idG} Let $S = S_{\mu_{1}\cdots\mu_{n}\nu \lambda}\in G^{n+1}(M)$,
for $n\geq 1$.
Then
\[ S_{\mu_{1}\cdots\mu_{n}\nu\lambda} = 
\frac{n+1}{n+2}\left( S_{ [ \mu_{1}\cdots\mu_{n}\nu ] \lambda}
+ S_{ [ \mu_{1}\cdots \mu_{n}\lambda ] \nu} \right) \]
\end{lemma}
{\it Proof:}\\
For notational simplicity we suppress the $\mu\hspace{0.01in}$s and write
$S=S_{1\cdots n \nu\lambda}$. Let $\tau \in S_{n}$ be the cyclic permutation
$(1\hspace{0.07in} 2\hspace{0.07in} 3 \hspace{0.07in} \cdots \hspace{0.07in} n)$.
Then
\[2nS_{1\cdots n \nu\lambda} =
\sum_{i=0}^{n-1} (-1)^{i(n+1)} \left( 
    S_{\tau^{i}(1)\cdots\tau^{i}(n)\nu\lambda}
    + S_{\tau^{i}(1)\cdots\tau^{i}(n)\lambda\nu}  \right) \]
Property (\ref{G}) for elements of $G^{n+1}$ takes the form
\begin{eqnarray} 
S_{\tau^{i}(1)\cdots \tau^{i}(n) \nu\lambda}  & = & 
 (-1)^{n} S_{\tau^{i}(2)\cdots \tau^{i}(n) \nu\lambda \tau^{i}(1)}  
  - S_{\tau^{i}(3)\cdots \tau^{i}(n) \nu\lambda \tau^{i}(1) \tau^{i}(2)} + \nonumber \\
 &  & (-1)^{n} S_{\tau^{i}(4)\cdots \tau^{i}(n) \nu\lambda \tau^{i}(1) \tau^{i}(2)\tau^{i}(3)}  
 -   \cdots  + (-1)^{n} S_{\lambda \tau^{i}(1)\cdots \tau^{i}(n) \nu}   
\end{eqnarray}
Similarly, with the indices $\nu$ and $\lambda$ switched,
\begin{eqnarray} 
S_{\tau^{i}(1)\cdots \tau^{i}(n) \lambda \nu}  & = & 
 (-1)^{n} S_{\tau^{i}(2)\cdots \tau^{i}(n) \lambda \nu \tau^{i}(1)}  
  - S_{\tau^{i}(3)\cdots \tau^{i}(n) \lambda \nu \tau^{i}(1) \tau^{i}(2)} + \nonumber \\
 &  & (-1)^{n} S_{\tau^{i}(4)\cdots \tau^{i}(n) \lambda  \nu \tau^{i}(1)\tau^{i}(2)\tau^{i}(3)}  
 -   \cdots  + (-1)^{n} S_{\nu \tau^{i}(1)\cdots \tau^{i}(n) \lambda}   
\end{eqnarray}
In the sum $ S_{\tau^{i}(1)\cdots\tau^{i}(n)\nu\lambda}
    + S_{\tau^{i}(1)\cdots\tau^{i}(n)\lambda\nu}$
all the middle terms on the right hand side of (6) and (7) cancel 
leaving only the end terms. Thus
\begin{eqnarray}
&   & 2n(-1)^{n}S_{1\cdots n \nu\lambda} \nonumber \\
& = & \sum_{i=0}^{n-1} (-1)^{i(n+1)} \left( 
S_{\nu \tau^{i}(1)\cdots \tau^{i}(n) \lambda} 
+ S_{\tau^{i}(2)\cdots \tau^{i}(n) \nu\lambda \tau^{i}(1)} \right)  
 + (\nu \leftrightarrow \lambda)  \nonumber \\
& = & \sum_{i=0}^{n-1} (-1)^{i(n+1)} \left( 
S_{\nu \tau^{i}(1)\cdots \tau^{i}(n) \lambda} +(-1)^{n+1}
S_{\nu\tau^{i}(2)\cdots \tau^{i}(n)  \tau^{i}(1)\lambda} \right)  
 +  (\nu \leftrightarrow \lambda)  \nonumber \\
& = &  \sum_{i=0}^{n-1} (-1)^{i(n+1)}  
S_{\nu \tau^{i}(1)\cdots \tau^{i}(n) \lambda} +
\sum_{i=0}^{n-1} (-1)^{(i+1)(n+1)}
S_{\nu\tau^{i}(2)\cdots \tau^{i}(n)  \tau^{i}(1)\lambda}   
 + (\nu \leftrightarrow \lambda)   \nonumber \\
& = & 2 \sum_{i=0}^{n-1} (-1)^{i(n+1)}  
S_{\nu \tau^{i}(1)\cdots \tau^{i}(n) \lambda} 
+  (\nu \leftrightarrow \lambda)   \nonumber 
\end{eqnarray}
Dividing by $2$ and then adding $2(-1)^{n}S_{1\cdots n \nu\lambda}$ to both sides gives
\begin{eqnarray}
&   & (n+2)(-1)^{n}S_{1\cdots n \nu\lambda} \nonumber \\  
& = & (-1)^{n}S_{1\cdots n \nu\lambda}+ \sum_{i=0}^{n-1} (-1)^{i(n+1)}  
S_{\nu \tau^{i}(1)\cdots \tau^{i}(n) \lambda} 
+  (\nu \leftrightarrow \lambda)   \nonumber  \\
& = &  S_{\nu 1\cdots n \lambda} + (-1)^{n}S_{1\cdots n \nu\lambda}
+ (-1)^{n+1} S_{\nu 2\cdots n 1 \lambda} 
+ S_{\nu 3\cdots n 1 2\lambda} +  \nonumber \\
& & (-1)^{n+1}S_{\nu 4 \cdots n 1 2 3\lambda} + \cdots + 
(-1)^{n+1} S_{\nu n 1 2\cdots n-1 \lambda} 
 + (\nu \leftrightarrow \lambda)   \nonumber \\
& = &  S_{\nu 1\cdots n \lambda} + (-1)^{n}S_{1\cdots n \nu\lambda}
+ S_{ 2\cdots n \nu 1 \lambda} 
+ (-1)^{n} S_{3\cdots n \nu 1 2\lambda} + \nonumber \\
& & S_{ 4 \cdots n\nu 1 2 3\lambda} + \cdots + (-1)^{n}S_{ n \nu 1 2\cdots n-1 \lambda} 
 + (\nu \leftrightarrow \lambda)   \nonumber 
\end{eqnarray}
Consider the identity
\begin{eqnarray}
& & S_{[1\cdots n \nu ]\lambda} \nonumber \\
& = & \frac{1}{n+1} \left(
S_{[1\cdots n] \nu\lambda}
+ (-1)^{n}S_{[ 2\cdots n \nu ]1 \lambda} 
+ \cdots + (-1)^{n}S_{[\nu 1\cdots n-1] n \lambda} \right) 
\nonumber \\
& = & \frac{1}{n+1} \left(
S_{1\cdots n \nu\lambda}
+ (-1)^{n}S_{ 2\cdots n \nu 1 \lambda} 
+ \cdots + S_{ n \nu 1 2\cdots n-1 \lambda}+ 
(-1)^{n}S_{\nu 1\cdots n \lambda} \right) 
\nonumber \\
& = & \frac{(-1)^{n}}{n+1} \left(
S_{\nu 1\cdots n \lambda} + (-1)^{n}S_{1\cdots n \nu\lambda}
+ S_{ 2\cdots n \nu 1 \lambda} 
+ \cdots + (-1)^{n}S_{ n \nu 1 2\cdots n-1 \lambda} \right) \nonumber
\end{eqnarray}
Substitute this into the above expression for 
$(n+2)(-1)^{n}S_{1\cdots n \nu\lambda}$ to obtain
\[ S_{1\cdots n \nu\lambda} = 
\frac{n+1}{n+2}\left( S_{ [ 1 \cdots n \nu ] \lambda}
+ S_{ [ 1 \cdots n\lambda ] \nu} \right)\]
$\Box$ \\

\begin{proposition} 
$ \phi\circ\psi = Id_{G^{*}} $
\end{proposition}
{\it Proof:}\\
Let $S = S_{\mu_{1}\cdots\mu_{n}\nu \lambda}\in G^{n+1}(M)$, for $n\geq 1$. Then
\begin{eqnarray}
\phi\circ\psi (S)_{\mu_{1}\cdots\mu_{n}\nu\lambda}  
& = & \frac{1}{2}\left( \psi(S)_{\mu_{1}\cdots\mu_{n}\nu \lambda}
                       + \psi(S)_{\mu_{1}\cdots\mu_{n} \lambda \nu} \right) \nonumber \\
& = & \frac{n+1}{n+2}\left( S_{ [ \mu_{1} \cdots \mu_{n} \nu ] \lambda} +  
  S_{ [ \mu_{1} \cdots \mu_{n} \lambda ] \nu} \right) \nonumber \nonumber \\
& = &  S_{\mu_{1}\cdots\mu_{n}\nu \lambda} \nonumber 
\end{eqnarray}
where the last equality follows from the lemma. \\
$\Box$ 

\begin{corollary}
 $\phi$ and $\psi$ are inverse maps and $K^{n}(M)$ is canonically isomorphic
 to $G^{n}(M)$ for all $n$.
\end{corollary}

We define $d_{G}:G^{*}(M)\longrightarrow G^{*}(M)$ by
\[  d_{G} = \phi \circ d_{K} \circ \psi \]
Equivalently, $d_{G}$ is the unique map that makes the following diagram
commutative

\begin{picture}(400,135)(-45,10)
\put(15,100){$\cdots$}
\put(39,103){\vector(1,0){30}}
\put(81,100){$K^{n}(M)$} \put(130,103){\vector(1,0){30}} \put(173,100){$K^{n+1}(M)$} 
\put(232,103){\vector(1,0){30}} \put(273,100){$\cdots$}

\put(15,44){$\cdots$}
\put(39,47){\vector(1,0){30}}
\put(81,44){$G^{n}(M)$} \put(130,47){\vector(1,0){30}} \put(173,44){$G^{n+1}(M)$} 
\put(232,47){\vector(1,0){30}} \put(273,44){$\cdots$}

\put(103,91){\vector(0,-1){30}} \put(97,61){\vector(0,1){30}}
\put(200,92){\vector(0,-1){30}} \put(194,62){\vector(0,1){30}}

\put(50,110){\footnotesize{$d_{K}$}} \put(50,33){\footnotesize{$d_{G}$}}
\put(142,110){\footnotesize{$d_{K}$}} \put(142,33){\footnotesize{$d_{G}$}}
\put(242,110){\footnotesize{$d_{K}$}} \put(242,33){\footnotesize{$d_{G}$}}

\put(84,72){\footnotesize{$\psi$}} \put(109,72){\footnotesize{$\phi$}}
\put(181,72){\footnotesize{$\psi$}} \put(206,72){\footnotesize{$\phi$}}

\end{picture} \\
In particular, if $S=S_{\mu \nu} \in G^{1}(M)$ then
\[ (d_{G}S)_{\mu \nu \lambda} = \frac{1}{2}\nabla_{\mu}S_{\nu \lambda}
 -\frac{1}{4}\left( \nabla_{\nu}S_{\mu \lambda} + \nabla_{\lambda}S_{\mu \nu} \right) \] 
 
$(K^{*}(M),d_{K})$ and $(G^{*}(M),d_{G})$ are naturally isomorphic cochain
complexes with inverse cochain maps $\phi$ and $\psi$. A trivial consequence is
\begin{theorem}
$ H_{G}^{*}(M) \cong H_{K}^{*}(M) $
\end{theorem}

\section{MP Spacetimes}

We now show that the symmetric-free part of the Levi-Civita connection 
of MP spacetimes and the Schwarzschild solution may be 
represented by means of a gravitational potential in the $(G^{*}(M),d_{G})$
complex. As mentioned in the introduction, since the Levi-Civita connection is not a tensor
this procedure is coordinate dependent and has the effect of selecting, within
the full diffeomorphism group, those coordinate systems for which a gravitational
potential exists, if in fact it does. The gravitational field is then viewed as a field 
propagating on a flat background structure defined by such a preferred coordinate system.

The MP spacetimes are a class of analytic solutions to the field equations of
general relativity in an electromagnetic field given by
\[ \begin{array}{lll}
ds^{2} & = & H^{-2}dt^{2} - H^{2}d\vec{x}^{2}, \\
A_{\mu} & = & \delta_{\mu t}\alpha(H^{-1} -1) 
\end{array} \]
where $\vec{x}=(x^{1},x^{2},x^{3})$, $\alpha = \pm 2$ and $H=H(\vec{x})$ is harmonic
in the $\vec{x}$ variables.
For a charge $q$, $\alpha = -2sign(q)$ and $H=1+\frac{G_{N}M}{|\vec{x}|}$ this gives
the Extreme Reissner-Nordstr\"{o}m black hole.

The Schwarzschild solution may also be expressed in isotropic coordinates as
\[ ds^{2}=\left(1+\frac{\omega/4}{\rho}\right)^{2}\left(1-\frac{\omega/4}{\rho}\right)^{-2}
   dt^{2}-\left(1-\frac{\omega/4}{\rho}\right)^{4}d\vec{x}^{2} \]
where 
\[ r = |\vec{x}| = 
             \frac{\left(\rho-\frac{\omega}{4}\right)^{2}}{\rho} \]

More generally, consider any metric in isotropic coordinates of the form
\[ ds^{2}=f(H)dt^{2}-g(H)d\vec{x}^{2} \]
with  Levi-Civita connection $\Gamma^{\mu}_{\nu \lambda}$,
where $f$ and $g$ are arbitrary smooth functions of a single variable and $H=H(\vec{x})$.
We associate to the coordinates a metric 
\[\eta = diag(+1,-1,-1,-1)\]
with which indices are raised and lowered. 
Define the symmetric-free part $F_{\mu\nu\lambda}$ of 
$\Gamma_{\mu\nu\lambda} =  \eta_{\mu \tau}\Gamma^{\tau}_{\nu \lambda}$ by
\[ F_{\mu \nu \lambda} = \Gamma_{\mu \nu \lambda} - \Gamma_{(\mu \nu \lambda)}  \]
where  
\[ \Gamma_{(\mu_{1} \mu_{2} \mu_{3})} = \frac{1}{3!}
   \sum_{\sigma \in S_{3}}\Gamma_{\mu_{\sigma(1)} \mu_{\sigma(2)} \mu_{\sigma(3)}} \]
Then $F_{(\mu\nu\lambda)} =0$ and since $F_{\mu\nu\lambda}$ is symmetric
in the $\nu,\lambda$ indices, 
\[ F_{\mu\nu\lambda} +F_{\nu\lambda\mu}+ F_{\lambda\mu\nu} =0 \]
This is property (\ref{G}) and so $F= F_{\mu\nu\lambda} \in G^{2}(M)$.

The non-zero Christoffel symbols for $ds^{2}=f(H)dt^{2}-g(H)d\vec{x}^{2}$ are 
\[ \begin{array}{ll}
\Gamma_{t j t} =  \Gamma_{t t j}  =  \frac{1}{2}\partial_{j}log f(H) 
& \hspace{0.5in} 1\leq j \leq 3 \nonumber \\
\Gamma_{j t t}  =  -(\partial_{j}f(H))/2g       
& \hspace{0.5in} 1\leq j \leq 3  \nonumber \\
\Gamma_{kjk} = \Gamma_{kkj}  =  -\frac{1}{2}\partial_{j}log g(H)  
& \hspace{0.5in} 1\leq j,k \leq 3 \nonumber \\
\Gamma_{jkk}   =  \frac{1}{2}\partial_{j}log g(H)  
& \hspace{0.5in} 1\leq j \neq k \leq 3 \nonumber \\
\end{array} \] 
Here we have denoted $x^{j}$ and $x^{k}$ simply by $j$ and $k$, respectively. 

Let $u$ be a solution to 
\[ u'=-\frac{2f'}{3f}-\frac{2f'}{3g} \]
and set 
\[ v=\frac{4}{3}log g\]
The symmetric-free part
$F_{\mu \nu \lambda}$ of $\Gamma_{\mu \nu \lambda}$ is
\[ \begin{array}{ll}
F_{t j t} =  F_{t t j}  =  -\frac{1}{4}\partial_{j}u(H) 
& \hspace{0.5in} 1\leq j \leq 3 \nonumber \\
F_{j t t}  =  \frac{1}{2}\partial_{j}u(H)       
& \hspace{0.5in} 1\leq j \leq 3  \nonumber \\
F_{kjk} = F_{kkj}  =  -\frac{1}{4}\partial_{j}v(H)  
& \hspace{0.5in} 1\leq j\neq k \leq 3 \nonumber \\
F_{jkk}   =  \frac{1}{2}\partial_{j}v(H)  
& \hspace{0.5in} 1\leq j \neq k \leq 3 \nonumber \\
\end{array} \] 
and all other terms equal to zero.
Let $A = A_{\mu \nu} \in G^{1}(M)$ be defined by
\[ A = u(H)dt^{2}+v(H)d\vec{x}^{2} \]
It is straightforward to verify that 
\[ F=d_{G}A \] 
That is, $F$ is the coboundary of the gravitational potential $A$ in the $G^{*}(M)$ complex.

\end{document}